\documentclass[letterpaper,twocolumn,10pt]{article}
\usepackage{usenix,epsfig,endnotes}
\usepackage{amsmath, amsthm, amssymb, amsfonts}
\usepackage{tikz}
\usepackage{amsmath}
\usepackage{filecontents}
\usepackage{caption}
\usepackage{subcaption}
\usepackage{xcolor}
\usepackage{comment}
\usepackage{breakcites}
\newcommand{\nearopt}{\textsf{Near-Opt}}
\begin{document}
\title{Characterizing TCP's Performance for Low-Priority Flows Inside a Cloud}
\author{
{\rm Hafiz Mohsin Bashir}\\
Tufts University
\and
{\rm Abdullah Bin Faisal}\\
Tufts University
\and
{\rm Fahad R. Dogar}\\
Tufts University
}
\maketitle
\begin{abstract}
Many cloud systems utilize low-priority flows to achieve various performance objectives (e.g., low latency, high utilization), relying on TCP as their preferred transport protocol. However, the suitability of TCP for such low-priority flows -- specifically, how \emph{prioritization-induced} delays in packet transmission can cause spurious timeouts and low utilization -- is relatively unexplored.  In this paper, we conduct an empirical study to investigate the performance of TCP for low-priority flows under a wide range of realistic scenarios: use-cases (with accompanying workloads) where the performance of low-priority flows is crucial to the functioning of the overall system as well as various network loads and other network parameters. Our findings yield two key insights: 1) for several popular use-cases (e.g., network scheduling), TCP's performance for low-priority flows is within $2\times$ of a near-optimal scheme, 2) for emerging workloads that exhibit an \emph{on-off} behavior in the high priority queue (e.g., distributed ML model training), TCP's performance for low-priority flows is poor. Finally, we discuss and conduct preliminary evaluation to show that two simple strategies -- weighted fair queuing (WFQ) and \emph{cross-queue} congestion notification -- can substantially improve TCP's performance for low-priority flows. 
\end{abstract}
\section{Introduction}
\label{sec:intro}

In this measurement paper, we investigate the interplay of two important aspects of today's cloud systems: (1) the widespread adoption of network prioritization and (2) the prevalent use of TCP as the primary transport protocol.
Our specific focus is on understanding the suitability of TCP for low-priority flows, and we frame our inquiry as: \textit{Is TCP good enough for low-priority flows under a cloud setting?}

Modern cloud systems host a multitude of applications with varying performance objectives.
To support these needs, there has been a widespread move towards adopting network prioritization (i.e., using priority queues inside switches). 
Network prioritization guarantees isolation across different types of applications, enabling various use-cases such as network flow scheduling (e.g., DAS~\cite{das}, pFabric~\cite{pfabric}, Homa~\cite{homa}, Baraat~\cite{baraat}) and workload co-location (e.g., perfIso~\cite{perfIso}, Heracles~\cite{heracles}, SmartHarvest~\cite{smart-harvest}).
For example, EBB~\cite{ebb}, Meta's backbone network, services several traffic classes in a strictly prioritized manner.
A new use-case of prioritization is in the context of RDMA-based stacks where network operators can use the TCP/IP traffic as a reliable and robust low-priority backup, with RDMA-based traffic as the higher priority traffic in the system~\cite{pangu}.
For many of the above use-cases, the performance of low-priority flows -- albeit less important than the high priority traffic -- is still important. 
For example, in systems like DAS~\cite{das} which uses duplication to reduce tail latency, a lower priority flow may determine the application performance if the corresponding high priority flow becomes a straggler. 

On the other hand, despite a large body of work highlighting the performance implications of using TCP for latency-critical applications~\cite{pdq,p3,pfabric,d2tcp,dctcp}, it remains the defacto protocol in today's cloud systems for important reasons such as backward compatibility and stability~\cite{morgan-stanley-nsdi15,accelnet,pangu,luna}. 
For instance, a recent work from Alibaba~\cite{luna} emphasized that their preference for TCP stems from the lack of specialized in-network support and the presence of a fleet of servers lacking support for emerging technologies (e.g., DPDK).

This motivates an investigation into potential issues that may arise when TCP is used for low-priority flows under network prioritization.
To the best of our knowledge, there has been no prior research dedicated to investigating this specific inquiry.
Intuitively, the use of TCP for low-priority flows can introduce a range of issues. This is primarily because low-priority flows are susceptible to extended queuing delays, which, in the worst-case scenario, can be unbounded~\cite{conquest}. These delays impact TCP's feedback loop, potentially resulting in unnecessary backoffs, and spurious retransmissions, where packets are mistakenly assumed to be lost while they are stuck within network switches.

Building on this intuition, our study evaluates the performance of TCP for low-priority flows across various important use-cases, encompassing a wide range of settings, including characteristics of high and low-priority traffic, the load and workload profiles, and specific TCP and switch configurations (e.g., RTO\textsubscript{min}, buffer size, etc.). 
Our study has two primary objectives: First, to shed light on the cases where TCP is a suitable choice for low-priority flows and identify scenarios where it is not.
Second, to explore minor adjustments that can help address any limitations observed in TCP's use for low-priority flows. Our methodology involves conducting experiments on a small-scale testbed and NS3 simulations~\cite{ns3}. We use a subset of experiments to cross-validate the results on both these platforms and then use simulations to explore a wide range of scenarios.

To quantify the impact of network prioritization and to uncover the room for improving performance of low-priority flows beyond what TCP offers, we consider a \emph{near-optimal} (\nearopt{}) fair transport protocol as a baseline. The \nearopt{} protocol has the advantage of instantaneous access to network state information, including available bandwidth, queue lengths, and packet loss details. 

Consequently, it represents an upper-bound that an efficient \emph{fair} transport protocol can achieve for low-priority flows. 
Our key measurement insights are as follows:
\begin{itemize}
    \item For several use-cases (e.g., network scheduling), TCP's performance for low-priority flows is within $2\times{}$ of \nearopt{}. This is because, for such use-cases, low-priority flows frequently get some fraction of network share allowing TCP to continually adapt to the network conditions.

    \item Low-priority TCP flows suffer the most --- more than $12\times{}$ compared to \nearopt{} --- when they coexist with high priority traffic that exhibits an \emph{on-off} behavior (typical of distributed ML model training workloads~\cite{p3,tiresias}). The \emph{on} period can completely starve the low-priority queue, causing TCP's feedback loop to stall and experience severe performance degradation. During the \emph{off} period, unnecessary congestion backoffs limit TCP from efficiently utilizing the link bandwidth. 

    \item Emerging technology trends such as increased link speeds and relatively smaller switch buffer sizes~\cite{one-more-config} significantly affect TCP's performance. Specifically when available buffer is equally divided between all the queues, TCP's performance is degraded by up to $5\times{}$ compared to when the entire buffer is available for the low-priority queue. 

    \item Two simple strategies -- weighted fair queuing (WFQ), and \emph{cross-queue} congestion notification -- effectively alleviate stalls in TCP's feedback loop, resulting in up to 9$\times$ reduction in completion times for low-priority flows.
        
\end{itemize}

The paper is organized as follows: We first present popular and emerging use-cases of network prioritization and highlight their performance implications for TCP when used for low-priority flows (\S\ref{sec:motive}). In~\S\ref{sec:eval-main}, we measure TCP's performance for low-priority flows across popular use-cases.
In~\S\ref{sec:micro}, we investigate the role of several configurations crucial to the performance of TCP such as emerging technology trends (e.g., smaller buffer sizes) and important network and transport parameters (e.g., RTO\textsubscript{min}).
Finally, in \S\ref{sec:future}, we discuss two simple strategies (and do the preliminary evaluation) to show substantial improvement in TCP's performance for low-priority flows.
\section{Preliminaries and Motivation}
\label{sec:motive}

We begin by highlighting important use-cases of network prioritization in modern cloud systems, emphasizing on the importance of low-priority traffic~(\S\ref{sec:priority-sched}).
We then discuss the performance implications of using TCP for low-priority flows~(\S\ref{sec:fairtrans-priosched}).

\subsection{Common Use-Cases of Network Prioritization}
\label{sec:priority-sched}
    
\paragraph{\textbf{Network Scheduling:}}
    Network prioritization is a cornerstone in the design of network scheduling schemes (e.g., flow/co-flow/task scheduling)~\cite{baraat,d3,pdq,das,pase,pias,pfabric,d2tcp}.
    For example, PIAS~\cite{pias} and pFabric~\cite{pfabric} use priority queues at network switches to minimize average flow completion times (FCTs). Similarly, duplication based schemes~\cite{das,primary-first,low-latency} leverage low-priority flows to minimize tail latency.
    For example, DAS~\cite{das} duplicates incoming request at a lower priority to mask \emph{stragglers}\footnote{Uncertain performance issues that commonly arise in large scale systems. These are notoriously hard to predict or mitigate at run time~\cite{ledge}} that contribute significantly to the tail latency. 
    Recently, a growing body of research highlights the benefits of incorporating network prioritization in the distributed training of machine learning models~\cite{p3,tictac,ccml-hotnets,bytesched}. This line of work further highlights the importance and relevance of this particular use-case. 

\paragraph{\textbf{Workload Co-location:}}
Cloud systems, particularly those hosting user-facing applications such as web search, are typically designed to handle peak loads. 
However, during the periods of low demand, this over-provisioning results in under-utilization of cloud resources (e.g., network, compute etc.,)~\cite{low-util-google-borg,low-util-google,low-util-twitter-quasar,low-util-dc-comp,wl-col-ali-elsticity}.
To improve the utilization of such systems and to reduce the total cost of ownership (TCO), several proposals advocate to co-locate best-effort workloads alongside latency-critical applications~\cite{heracles,bistro,wl-col-ali-elsticity,wl-col-ms-harvest,perfIso}. Co-location enables best-effort workloads to harvest spare system resources.
To ensure that the performance of latency critical applications is not negatively affected by the best-effort workloads, operators employ prioritization (i.e., giving latency critical workloads priority over best effort workloads) at potential bottleneck resources (e.g., network).
For example, \emph{PerfIso}~\cite{perfIso} leverages prioritization to safely scavenge spare system resources in \emph{Microsoft Bing's} cluster, enabling the operators to simultaneously maintain low \emph{response time} for latency critical services and high utilization of the overall system.

\paragraph{\textbf{Hybrid RDMA/TCP-IP Services:}}
A key requirement in a modern cloud system is the high availability of critical services (e.g., key-value stores) that provide support for many other applications.
The recent trend of cloud operators turning to RDMA~\cite{hpcc,dcqcn,timely,pangu,mp-rdma} has raised concerns about the availability of these critical services because of several outstanding challenges with the adoption of RDMA (e.g., pause frame storm)~\cite{collie,pangu,gentle-flow-control,deadlock-in-datacenter}.
An effective strategy is the joint-deployment of RDMA and TCP such that TCP runs in the background at very low load as a \textit{fail-safe} option.
Should RDMA experience an anomaly (e.g., PFC storm), TCP can take over, ensuring high availability~\cite{pangu,accelnet,brpc}.
For example, Gao et. al., integrated TCP and RDMA in Pangu~\cite{pangu} -- a cloud storage system -- and provided support for fine grained switching between RDMA and TCP to ensure high availability.
To keep TCP from interfering with RDMA, TCP traffic can be mapped to a lower priority queue at the switches.

\begin{table*}[]
\centering
\resizebox{0.75\textwidth}{!}{%
\begin{tabular}{|c|c|cc|c|}
\hline
\textbf{Use-Case} & \textbf{Objective(s)} & \multicolumn{2}{c|}{\textbf{\begin{tabular}[c]{@{}c@{}}Example\\ High / Low-Priority\end{tabular}}} & \textbf{Proposal(s)} \\ \hline
Scheduling & \begin{tabular}[c]{@{}c@{}}Policy Specific\\ (e.g., Improved FCTs)\end{tabular} & \multicolumn{1}{c|}{\begin{tabular}[c]{@{}c@{}}Small \\ Flows\end{tabular}} & \begin{tabular}[c]{@{}c@{}}Long \\ Flows\end{tabular} & \begin{tabular}[c]{@{}c@{}}PIAS\cite{pias}, Baraat\cite{baraat}, \\ DAS\cite{das}, pFabric\cite{pfabric}\end{tabular} \\ \hline
\begin{tabular}[c]{@{}c@{}}Workload \\ Co-location\end{tabular} & \begin{tabular}[c]{@{}c@{}}Increased\\ Utilization\end{tabular} & \multicolumn{1}{c|}{\begin{tabular}[c]{@{}c@{}}Latency\\ Critical\end{tabular}} & \begin{tabular}[c]{@{}c@{}}Batched \\ Workload\end{tabular} & \begin{tabular}[c]{@{}c@{}}perfIso\cite{perfIso}, Hercales\cite{heracles}, \\ Bistro\cite{bistro}, SmartHarvest\cite{smart-harvest}\end{tabular} \\ \hline
\begin{tabular}[c]{@{}c@{}}Hybrid\\ Services\end{tabular} & \begin{tabular}[c]{@{}c@{}}High \\ Availability\end{tabular} & \multicolumn{1}{c|}{RDMA} & TCP & Pangu\cite{pangu} \\ \hline
\end{tabular}%
}
\caption{ Synthesizes important use-cases of network prioritization and highlights that the performance of low-priority flows can be crucial in achieving a certain objective in a cloud system.}
\label{tab:usecases}
\end{table*}

\paragraph{\textbf{Synthesis:}} Table~\ref{tab:usecases} summarizes the above use-cases, highlighting how network prioritization helps achieve a diverse set of objectives by providing isolation across various abstractions (i.e., flows, workloads, services). Despite the pivotal role of low-priority flows in the success of these use-cases, their performance is generally overlooked in the existing work. In section~\S\ref{sec:eval-main}, we show that TCP's performance for low-priority flows varies across each of these use-cases.

\begin{figure}
  \centering
    \subfloat[Assumed network view]{\includegraphics[width=0.47\columnwidth]{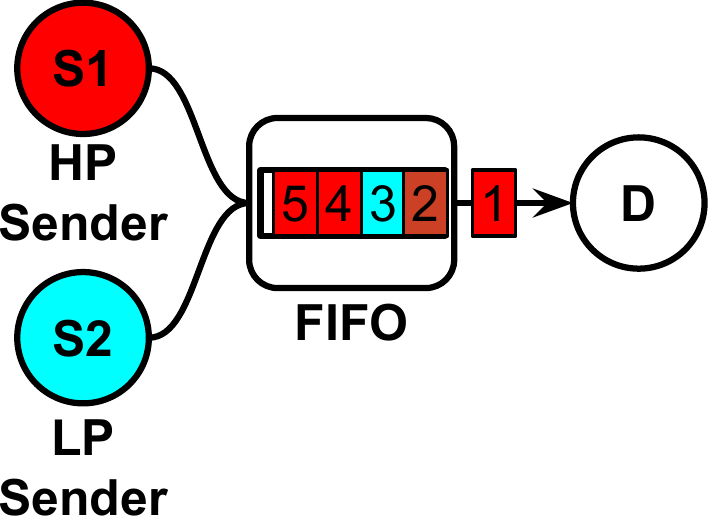}
    \label{fig:fairview}}
    \subfloat[Actual network view]{\includegraphics[width=0.47\columnwidth]{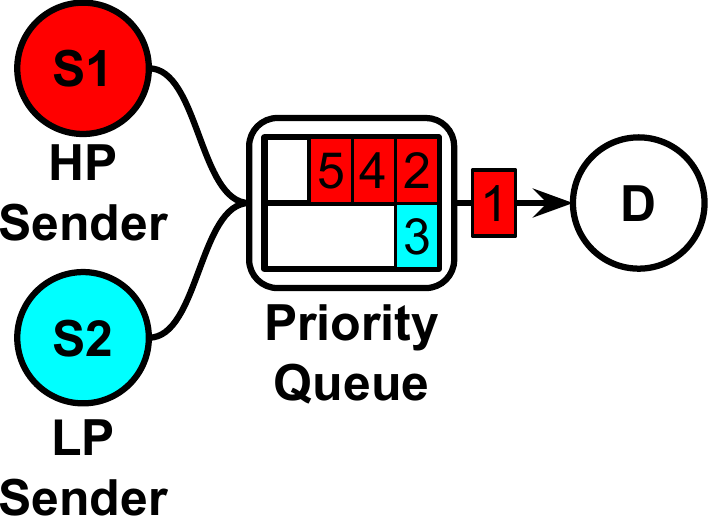}
    \label{fig:realview}}
    \caption{TCP assumes fair-scheduling inside the network. This assumption fails under priority scheduling. (a) depicts TCP's view of the network (i.e., all packets from low-priority (Blue) and high priority (Red) senders share the same queue). (b) shows that in reality packets from different priority classes are stored in their respective priority queue and they are serviced in the order of their priority.}
    \label{fig:diff-in-view}
\end{figure}

\subsection{TCP Under Network  Prioritization}
\label{sec:fairtrans-priosched}
    \textbf{Fair-share assumption of TCP:} TCP's congestion control assumes that data packets receive a ``fair'' (non-discriminatory) treatment inside the network (i.e., packets are assumed to be serviced in the order of their arrival). In reality, under network prioritization, the priority of a packet can override the service order.
    Figure~\ref{fig:diff-in-view} depicts that the difference between TCP's assumed view of the network~(\ref{fig:fairview}) and the actual view~(\ref{fig:realview}) results in longer queuing delays for low-priority packets -- unbounded in the worst case~\cite{conquest}. For example, the low-priority packet (shown in \emph{blue}) can only get the service once all the high priority packets (shown in \emph{red}) have been serviced.

\begin{figure}
  \centering
    \subfloat[High Retransmission]{\includegraphics[width=0.47\columnwidth]{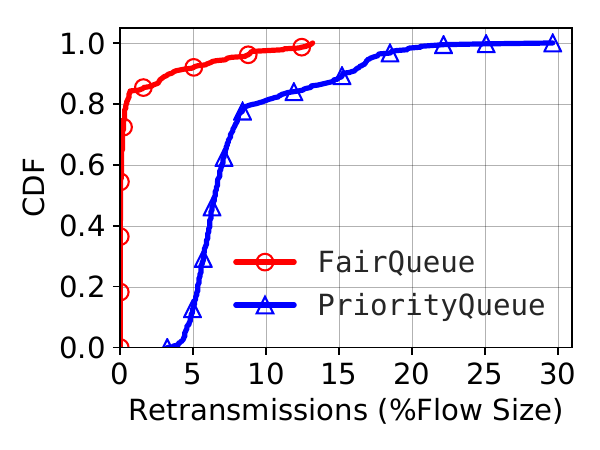}
    \label{fig:motive-retx}}
    \subfloat[High FCT]{\includegraphics[width=0.47\columnwidth]{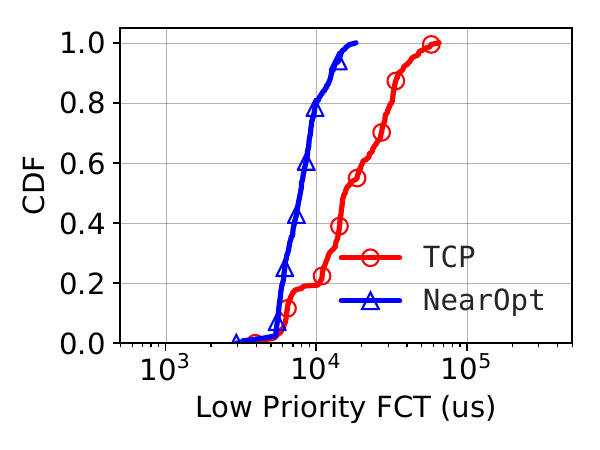}
    \label{fig:motive-fct}}
    \caption{Shows substantial impact of priority queuing on TCP's performance for low-priority flows. (a) highlights a significant increase in the retransmission rate experienced by low-priority flows under priority queues compared to fairshare. (b) highlights the impact of priority queuing on flow completion times (FCT) of low-priority flows between TCP and \nearopt{} (\S\ref{sec:eval-main}).}
    \label{fig:motive}
\end{figure}

\paragraph{\textbf{Implications for low-priority Flows:}} 
    TCP's congestion control relies on an end-to-end feedback loop (\emph{ack-clocking}) to estimate the available network bandwidth and to take congestion control decisions.
    Under network prioritization, high priority flows can delay the transmission of low-priority packets.
    This causes \emph{long} and highly \emph{variable} stalls in TCP's feedback loop which can be detrimental to the performance of low-priority flows.
    For example, the increased queuing delay can trigger timeouts which result in spurious retransmissions and unnecessary congestion back-offs.

    To quantify these observations, we conduct a small experiment on a 9 node cluster. We co-locate a background storage workload with a latency-critical workload. The latency-critical workload \emph{emulates} the network footprint of distributed ML training~\cite{p3,tiresias} and periodically generates a 8:1 incast that exhibits an \emph{on-off} communication pattern (more details in~\S\ref{sec:eval-wl-col}).
    Both applications share a common bottleneck link.
    We run this experiment with and without network prioritization.
    In the former case, flows from the storage workload are mapped to a lower priority queue compared to the flows from latency-critical application, whereas in the latter case, flows from both the applications share a single queue.

    Figure~\ref{fig:motive-retx} shows that the introduction of priority queues significantly increases the retransmission rate for the long flows (\S\ref{sec:eval-main}).
    For example, at the 80\textsuperscript{th} percentile (p80), the retransmission rate is approximately 8\% when network prioritization is enabled (PriorityQueue) while it is negligible in the case of no prioritization (FairShare). 
    This happens because of two main reasons:

\textbf{Spurious Timeouts:}
    Prioritization results in an increased queuing delay for low-priority flows. Consequently, TCP faces challenges in distinguishing between \emph{prioritization-induced} (caused by network prioritization) and the actual packet losses, which makes it susceptible to experiencing spurious timeouts (i.e., an in-flight packet is mistakenly assumed to be lost). TCP's current approach of estimating the timeout value as a long term moving average of round trip times proves limiting as it fails to adapt to \emph{transient} spikes in the round trip time caused by large bursts of high priority packets.
    
    \textbf{Convergence issues:}
    Bursty arrivals of high priority packets create large fluctuation in the available bandwidth for low-priority flows which leads to the following challenges:
    \begin{itemize}
        \item \emph{Overshoot:} 
        The presence of high priority traffic results in long feedback delays for low-priority flows. This slows down TCP's reaction to congestion and leads to the buildup of low-priority queues, causing a large number of packet losses.
        \item \emph{Undershoot:} The absence of high priority traffic yields the full link capacity to the low-priority flows. However, the congestion backoffs in the previous epoch of low bandwidth may result in an under-utilization of the network links.
    \end{itemize}

    To assess the impact of the above discussed issues on flow completion times, we repeat the experiment and substitute TCP with a \nearopt{} scheme which is designed to estimate near-optimal flow completion times of low-priority flows (\S\ref{sec:eval-main}).
    Figure~\ref{fig:motive-fct} captures the flow completion times of low-priority long flows. It clearly highlights the substantial opportunity to enhance TCP's performance. 

    While slow convergence is a well researched problem for TCP, the frequent stalls in the feedback loop coupled with high fluctuation in the available network bandwidth exacerbates this issue for low-priority flows.

\paragraph{\textbf{Conclusion:}}
The above discussed use-cases underscore the importance of low-priority flows in achieving a wide range of objectives in modern cloud systems (e.g., mitigating tail latency, ensuring high availability etc.,).
TCP's performance for low-priority flows, however, can be adversely affected by challenges such as spurious timeouts and convergence issues. 
This motivates the need to systematically study TCP's performance for low-priority flows under a diverse set of conditions.
\section{A Closer Look at TCP's Performance for Low-Priority Flows}
\label{sec:eval-main}

We now study TCP's performance for low-priority flows under a broad spectrum of use-cases~(\S\ref{sec:fairtrans-priosched}), load distribution across priority classes, choice of transport protocol for high priority traffic, and workloads.
Table~\ref{tab:eval-summary} synthesizes various settings across high and low-priority classes (e.g., load, flow size distribution) and highlights our key insights. 

We conduct experiments on both a small scale testbed and the NS3 simulator. The testbed experiments establish the credibility of the key insights of our evaluation, under realistic settings, with a real end-host stack and a network switch with prioritization support. It also helps us cross validate the simulator setup, which we use to explore a wider range of network and transport parameters. We present the testbed results in \S\ref{subsec:testbed-eval}; all others results are from our NS3 evaluation.

\begin{table*}[]
\centering
\resizebox{0.85\textwidth}{!}{%
\begin{tabular}{|c|c|c|l|}
\hline
\textbf{Use-Case} &
  \textbf{\begin{tabular}[c]{@{}c@{}}Load \\ Distribution\end{tabular}} &
  \textbf{\begin{tabular}[c]{@{}c@{}}Flow Size \\ Distribution\end{tabular}} &
  \multicolumn{1}{c|}{\textbf{Key Insights}} \\ \hline
\begin{tabular}[c]{@{}c@{}}Duplicate-Aware\\Scheduling (DAS)~\cite{das}\end{tabular} &
  Same &
  Same &
  \begin{tabular}[c]{@{}l@{}}Equal load across queues yields ample bandwidth for \\ low-priority flows to make progress. Consequently,\\ TCP does not suffer from issues discussed in~\ref{sec:motive}\end{tabular} \\ \hline
\begin{tabular}[c]{@{}c@{}}Shortest-Job-First\\ (SJF) \end{tabular} &
  Different &
  Different &
  \begin{tabular}[c]{@{}l@{}}The fraction of load induced by high priority flows determines\\ the impact of prioritization. The skewness in a workload's flow \\ size distribution plays a crucial role: greater skewness results\\  in higher performance penalty.\end{tabular} \\ \hline
\begin{tabular}[c]{@{}c@{}}Workload Colocation\end{tabular} &
  Different &
  Different &
  \begin{tabular}[c]{@{}l@{}}\emph{On-Off} network behavior of high priority workload\\ severely affects low-priority flows. This is primarily due to \\ the rapid fluctuations between very high and very low load, \\ which leads to high retransmissions and frequent backoffs.\end{tabular} \\ \hline
\begin{tabular}[c]{@{}c@{}}Hybrid Services\end{tabular} &
  Different &
  Same &
  \begin{tabular}[c]{@{}l@{}}The choice of high priority transport protocol has little impact.\\ Rather, the choice of workload is a dominant factor.\end{tabular} \\ \hline
\end{tabular}%
}
\caption{Compares high and low priority traffic classes in our evaluation scenarios across different axes and highlights key insights derived from each use-case. For example, in the context of DAS, both the high and low priority traffic experience the same load, and service the same workload (i.e., flow size distribution).}
\label{tab:eval-summary}
\end{table*}

\paragraph{\textbf{Experimental Setup:}}
For the NS3~\cite{ns3} evaluation, we use the following settings as default unless specified otherwise. We assume a single cluster setup where 40 nodes are interconnected via 10Gbps link to a single switch. We provision strict priority queues within both the end hosts and the network, comprising a high priority queue and a low-priority queue. 
The switch buffer is statically shared among ports. According to the settings discussed in an earlier work~\cite{one-more-config}, each switch port is allocated a 192KB buffer size, which is further partitioned in a 2:1 ratio between high and low-priority queues, respectively. In the absence of any traffic between servers, the average round trip time (RTT) is estimated to be 100us. 

We configure TCP-Cubic as the default congestion control algorithm and enable the selective acknowledgment (SACK) option. The initial congestion window is set to 2 packets, and the minimum retransmission timeout value (RTO\textsubscript{min}) is configured to 1ms. Additionally, we allocate 1MB for the send and receive buffers for TCP. To prevent priority-inversion within the driver queue, we limit the driver queue to 100 packets. We observed maximum throughput for TCP (when used for low-priority flows) under these settings. 
Finally, to generate traffic between servers, we use the publicly available DAS setup~\cite{das-setup} and tailored it for our use-cases.
We consider three load settings in our evaluation, \emph{low}, \emph{medium}, and \emph{high} which refer to network loads of $<20\%$, $20-40\%$ and $>40\%$ respectively. Similarly, we classify flows as \emph{small}, \emph{medium} and \emph{long} which refer to flow sizes of $<50KB$, $50KB-1MB$, and $>1MB$ respectively. 

\paragraph{\textbf{Baseline and Evaluation Criteria:}}
To quantify the impact of network prioritization and to uncover the room for improvement for low-priority flows using TCP, we consider an RCP~\cite{rcp} style near-optimal fair transport scheme (\nearopt{}) for low-priority traffic as our baseline.
\nearopt{} relies on two important pieces of information to provide a strong baseline: \emph{Loss Information (LI)} and \emph{Rate Information (RI)}\footnote{Note that \nearopt{} is \emph{only} used for comparison and may not be practically realizable.}.
\begin{figure*}[ht]
    \centering
      \subfloat[DAS - Medium Load]{
        \includegraphics[width=0.32\textwidth]{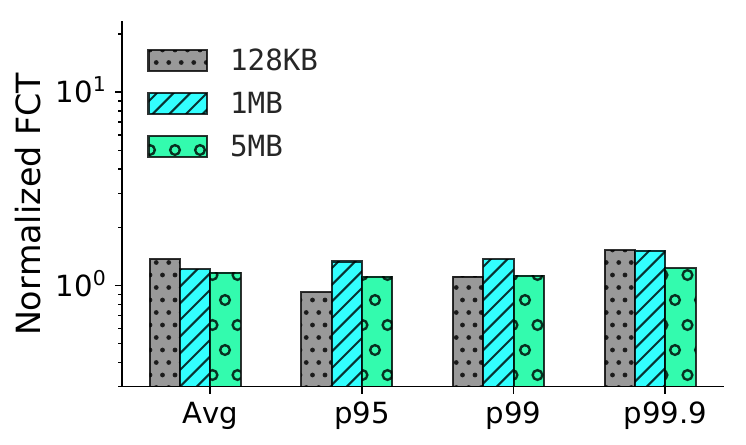}
        \label{fig:eval-netsched-das}}
    \subfloat[SJF - \emph{Web-Search}]{
        \includegraphics[width=0.32\textwidth]{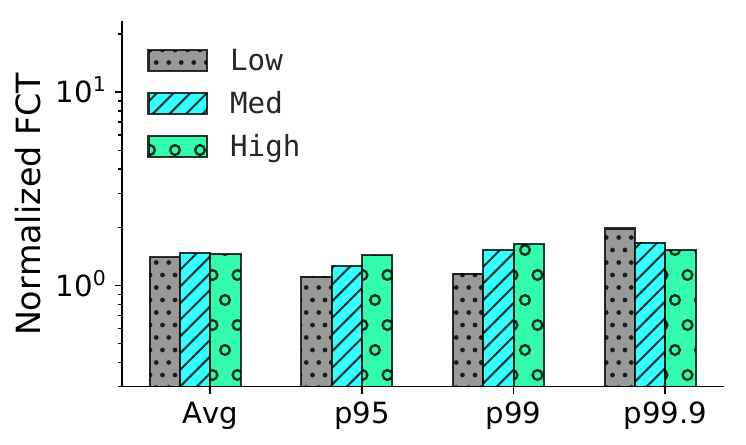}
        \label{fig:eval-netsched-sjf-web}}
    \subfloat[SJF - \emph{Data-Mining} ]{
        \includegraphics[width=0.32\textwidth]{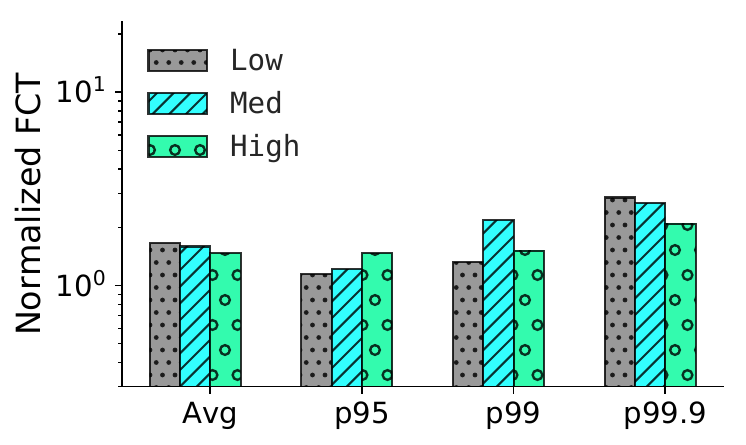}\label{fig:eval-netsched-sjf-fb}}
    \caption{TCP's performance for low-priority flows under network scheduling scenario. (a) Shows that for DAS~\cite{das} workload, TCP's performance for flows is insensitive to the flow sizes. (b) and (c) shows TCP's performance for long flows under shortest job first policy (SJF) for the \emph{web-search} and \emph{data-mining} workloads. Under SJF, we assign low-priority to long flows (size greater than 1 MB).}
\label{fig:eval-netsched}
\vspace{-0.1in}
\end{figure*}
\begin{itemize}
    \item \textbf{Loss Information:} Knowing whether a packet was lost in the network allows \nearopt{} to avoid spurious retransmissions. \nearopt{} makes use of a global data structure that records packet losses. When a timeout event occur, \nearopt{} looks up the global data structure to validate the loss allowing it to isolate spurious timeouts from true losses. The \nearopt{} can access this information in \textit{O(1)} time.
    
    \item \textbf{Rate Information:} By knowing the fair-share rate of a low-priority flow, \nearopt{} can avoid \emph{undershoot} and \emph{overshoot} issues~(\S\ref{sec:fairtrans-priosched}).
    \nearopt{} estimates the fair-share rate for each flow for the next RTT using the updated load and queue length information on a per link basis. 
    Specifically, it first calculates the high priority load by dividing the high priority bytes received in previous round (\emph{t-1}) --- sum of total bytes sent (\emph{B}) and the current queue length Q(\emph{t}) ---  by the round trip time (\emph{T}). 
    It then discounts the high priority load from the total link rate (\emph{L}) to estimate the available capacity for low-priority traffic and divides the result by the low-priority flow count on that link (\emph{F}) to keep track of the available flow rate\footnote{The rate estimation is done on the basis of load in the last round. The actual load in the next round can differ from the estimated load hence, \nearopt{} carries a small error. We choose \emph{T} as the link delay instead of RTT to minimize this error} on per link basis as shown in the equation below. Finally, it picks the minimum flow rate on all the links in the flow's path.
    \begin{eqnarray*}
        r_i(t) &=& \left(R - \sum_{c=0}^{i-1} B_c(t-1)/T - \sum_{c=0}^i Q_c(t)/T\right) \div |F|
    \end{eqnarray*}
    
\end{itemize}
Note that both the \emph{loss information} and parameters needed to compute a flow's rate are available to \nearopt{} in \emph{O(1)} time.

We use performance gap between TCP and \nearopt{} as our primary evaluation criteria. For this, we run each experiment using TCP and \nearopt{} for low-priority flows.
To highlight the performance gap, we normalize TCP's FCTs by the FCTs achieved using \nearopt{}. A lower (resp. higher) value of normalized FCT indicates superior (resp. poor) performance of TCP.
Since our focus is on the performance of low-priority flows, we only look at the FCTs of low-priority flows unless noted otherwise. 

We now evaluate TCP for three distinct use-cases (discussed in~\S\ref{sec:motive}) under a broad range of settings. 

\subsection{\textbf{Network Scheduling}}
\label{sec:eval-netsched}
We first look at how TCP performs for low-priority flows under two different network scheduling schemes:

\paragraph{\textbf{Duplicate Aware Scheduling (DAS)}}
DAS is a request duplication scheme which strives to minimize the tail latency of a cloud applications~\cite{das}.
DAS creates a duplicate copy of each incoming request at a lower priority and attempts to fetch both the responses (the primary and the duplicate) from two different servers. Upon retriving one copy, DAS purges the other one, a feature which we disable to simplify our experiment. 
A distinguishing property of this scenario is that both the workload and load across high and low-priority queues is the same.

\paragraph{Setup:} We consider a uniform flow size distribution with different mean values: 128KB, 1MB and 5MB.
We generate requests in a poisson process at medium load (30\%) which translates to a high aggregate load (60\% -- 30\% in each priority class) on the bottleneck link.

\paragraph{Observation:}
Figure~\ref{fig:eval-netsched-das} shows that under the DAS workload, TCP's performance is within 2$\times$ of \nearopt{}, across different flow sizes. 
This is because a relatively low load in the high priority queue provides ample bandwidth for low-priority flows to make progress. As a result, TCP is largely unaffected by the challenges posed by network prioritization.

\paragraph{\textbf{Shortest Job First (SJF)}}
Shortest job first (SJF) is well known to be optimal in minimizing the flow completion times~\cite{2D}.
Many popular schemes such as PIAS~\cite{pias}, PASE~\cite{pase}, pFabric~\cite{pfabric}, and PDQ~\cite{pdq} approximate this policy which makes it a good candidate in our evaluation~\footnote{Note that our goal here is not to argue about the efficacy of a scheduling policy, rather to investigate its performance implications on TCP for low-priority flows.}.
The key difference between our last experiment and this one lies in the variation of flow sizes and the distribution of load across priority queues.
\begin{figure*}[ht]
    \centering
    \subfloat[Across Load]{
        \includegraphics[width=0.32\textwidth]{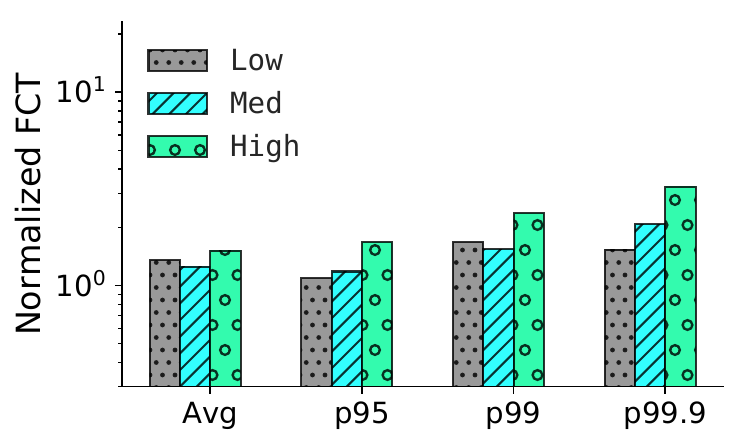}
        \label{fig:eval-wl-ml-load}}    
    \subfloat[Across flow sizes]{
        \includegraphics[width=0.32\textwidth]{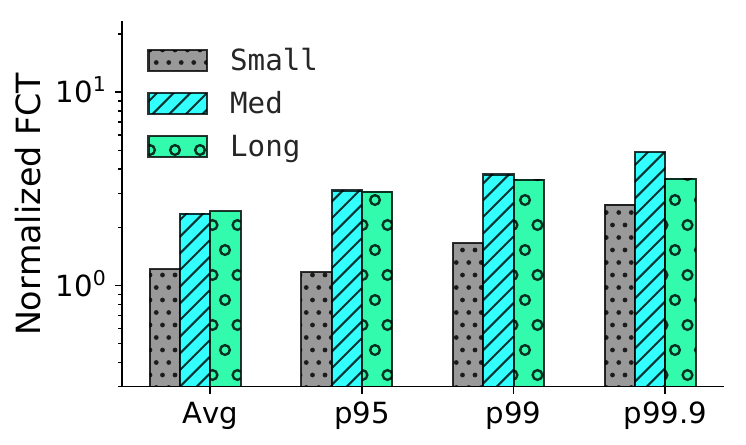}
        \label{fig:eval-wl-ml-fs}}
    \subfloat[Across Update Size]{
        \includegraphics[width=0.32\textwidth]{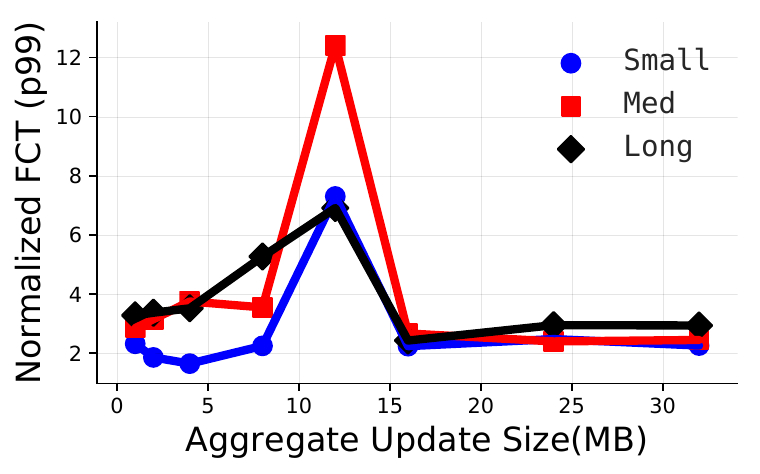}
        \label{fig:eval-wl-ml-ms}} 
    \caption{TCP's performance under workload co-location scenario across different loads, flow sizes and ML-model gradient update sizes in the high priority queue: 
    (a) shows the performance of TCP for low-priority flows when high priority traffic exhibit an \emph{on-off} behavior caused by the gradient update of size 4MB across different loads.
    (b) evaluates TCP across different flow size for the same workload at high load.
    (c) compares TCP's performance across different gradient update sizes.}
\label{fig:eval-wl-col}
\vspace{-0.1in}
\end{figure*}

\paragraph{Setup:}
For this experiment, we consider the \emph{web-search}~\cite{dctcp} and \emph{data-mining}~\cite{vl2} workloads. Following the guidelines in~\cite{dctcp}, we classify all flows of size greater or equal to 1MB as long flows.
We assign low-priority to the long flows and high priority to the rest of the flows (flow size less than 1MB). Flows are generated at \emph{low}, \emph{medium} and \emph{high} loads in a poisson process. 

\paragraph{Observation:} For the \emph{web-search} workload, the high priority flows are relatively large compared to the \emph{data mining} workload hence, for a given load setting, the flow arrivals in the high priority queue tend to be less bursty.
Contrarily, the \emph{data-mining} workload exhibits extreme skewness, with over 80\% of the flows being less than 10KB in size (hence, the arrival process tends to be bursty). Approx. 3\% of the flows are larger than 35MB and contribute to about 95\% to the overall load~\cite{pfabric}. 

Figure~\ref{fig:eval-netsched-sjf-web} shows that TCP's performance is within 2$\times$ to the \nearopt{} for web-search workload whereas, for the data-mining workload this difference can be up to 3$\times$. For instance, under high load, at 99.9\textsuperscript{th} percentile, the performance gap between TCP and \nearopt{} is 44\% larger for the data-mining workload compared to the \emph{web-search} workload. 

Overall, under both web-search and data-mining workloads, the low-priority queues experience the bulk of the total load (low load in the high priority queue).
Consequently, similar to the DAS experiment, this yields enough bandwidth for low-priority flows, shielding TCP from the challenges of network prioritization.

\subsection{\textbf{Workload Co-Location}}
\label{sec:eval-wl-col}

We now consider the scenario where two different applications are co-located to share cloud resources.
We broadly focus on the storage harvesting case -- a common scenario to utilize the spare storage capacity in compute intensive clusters~\cite{wl-col-ms-harvest}. 

\paragraph{Experiment Setup:}
We consider a storage workload from Alibaba's cluster~\cite{alibaba_storage} to generate low-priority flows. Since storage applications are often bottlenecked by the storage media, they tend to have a smaller network footprint; thus, we consider low load in the low-priority queue for this workload.
For high priority traffic, we consider a distributed machine learning (ML) training workload.
Distributed ML training typically involves utilizing a parameter server~\cite{psarch-osdi14} to periodically aggregate (disseminate) gradient updates from (to) multiple workers -- a process that takes place after each training iteration~\cite{tiresias} and often leads to an incast on the parameter server link, when updates from multiple workers converge. The combination of incast and the periodicity in the arrival process of these updates gives rise to an \emph{on-off} network behavior.
The specific properties of the \emph{on-off} behavior (e.g., the aggregate load, the size and frequency of \emph{on} periods) is a function of various factors such as the machine learning models and the underlying hardware and the software stack~\cite{tiresias,byteps,theory-distributed-dnn}. 
By default, we consider a gradient update size of 500KB from each worker (also referred as flow size), resulting in an aggregate update of 4MB (8:1 incast) on the parameter server's link. By adjusting the frequency of these gradient updates, we can control the load (low, medium, and high) on the bottleneck link. Additionally, we will assess the impact of different gradient update sizes in subsequent evaluations.

\paragraph{Observation:} 
Figure~\ref{fig:eval-wl-ml-load} exhibits TCP's performance across different load settings. Notably, TCP experience more than 4$\times$ performance difference compared to the \nearopt{} under high load while this gap is within 2$\times$ at low and medium loads. This outcome is expected because of the continued availability of link capacity for low-priority flows at low and medium loads which keeps the feedback loop of low-priority flows active.

Zooming in on the high load scenario, in figure~\ref{fig:eval-wl-ml-fs} we examine the performance implications across different flow sizes. We observe that TCP experience performance degradation regardless the flow size. The medium flows suffers the most, up-to $4.9\times$.
To further investigate this phenomenon, we repeat the experiment across different update sizes. 
Figure~\ref{fig:eval-wl-ml-ms} shows that as the size of gradient updates increases, the performance gap between TCP and \nearopt{} continues to widen. Specifically, when the update size reaches 12MB, the performance gap surpasses $10\times$. For updates of larger sizes, the long duration of \emph{on} period in the high priority queue results in the starvation of the low-priority flows. Consequently, the duration of \emph{on} period becomes a dominant factor in determining the flow completion times which masks the protocol's inefficiencies and makes the choice of protocol irrelevant resulting in a significant reduction in the performance gap between TCP and \nearopt{}.

\subsection{\textbf{Hybrid Services}}\label{subsubsec:eval-hybservice}
\label{sec:eval-hybserv}
We now consider the use-case of hybrid services.
Recall that a key aspect of this use-case is the co-existence of a high performance transport protocol (e.g., HPCC~\cite{hpcc}) for the high-priority traffic and TCP for low-priority traffic.
Unlike TCP -- which requires multiple round-trip times to fully utilize the link -- a high performance transport protocol can quickly saturate the link hence, the presence of a long flow in the high priority queue can completely halt the feedback loop of a low-priority flow.

\paragraph{Setup:}
For the high priority traffic, we use \nearopt{} as our high performance transport due to its ability to quickly saturate link bandwidth.
We use one node as a storage client while other nodes in the cluster act as storage servers.
The client fetches storage objects using two independent poisson processes (one for each priority class).
We use fixed flow sizes of 32KB, 64KB,128KB, and 1MB and keep the aggregate load (high priority+low-priority) at 80\% while changing its distribution across the two priority classes.

\begin{figure}
  \centering
    \subfloat[Impact of Flow Size]{\includegraphics[width=0.48\columnwidth]{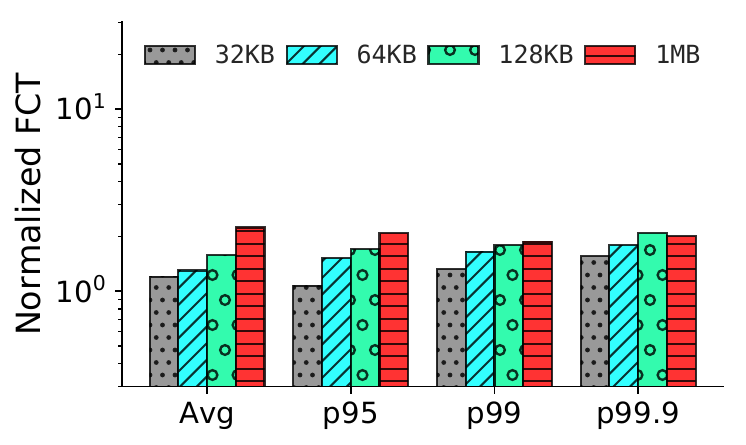}
    \label{fig:eval-hybservice-30P}}
    \subfloat[Impact of Load]{\includegraphics[width=0.48\columnwidth]{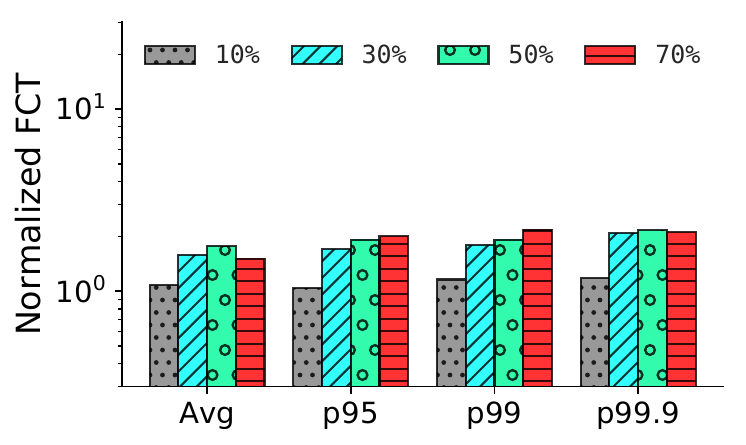}
    \label{fig:eval-hybservice-128KB}}    
	\caption{Shows TCP's performance for \textit{Hybrid Services} use-case at 80\% total load (High + low-priority). (a)  Shows TCP's performance at 30\% load in the low-priority queue across different flow sizes. In this case, TCP's performance improves as the flow size decreases. (b) Shows TCP's performance for 128KB flow sizes across different load settings for low-priority queue. With increased load in the low-priority queue the performance gap between TCP and \nearopt{} increases.}
    \label{fig:eval-hybservice}
\end{figure}

\paragraph{Observation:}
Fig~\ref{fig:eval-hybservice-30P} shows TCP's performance across different flow sizes at medium load in the low-priority queue (50\% load in high priority queue - 30\% load in the low-priority).
In this particular use-case, we observe that TCP exhibits superior performance for short flows compared to the long flows. This can be attributed to the high efficiency of \nearopt{} in the high priority queue, which can rapidly saturate the link, consequently having a greater impact on the feedback loop of low-priority flows. As long flows remain in the network for an extended duration, they become more susceptible to repeated stalls, leading to performance issues.
Fig~\ref{fig:eval-hybservice-128KB} shows TCP's performance across different loads in the low-priority queue. We notice that the load settings matter less and TCP experience similar performance across different load configurations.
In conclusion, for this use-case, we find that the flow size (workload) is the dominant factor in determining the performance implications for TCP.

\begin{figure}[!ht]
    \begin{minipage}{\linewidth}
        \begin{subfigure}{.32\linewidth}
            \includegraphics[width=\linewidth]{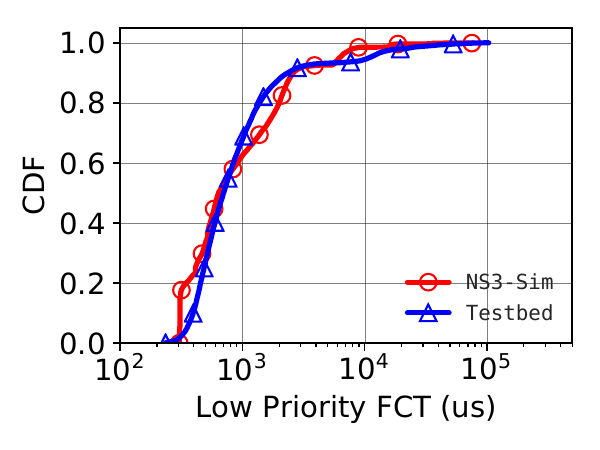}
            \caption{\emph{On-Off} - Small}
            \label{fig:TBVSSim-WLCOL-Small}
        \end{subfigure}%
        \begin{subfigure}{.32\linewidth}
            \includegraphics[width=\linewidth]{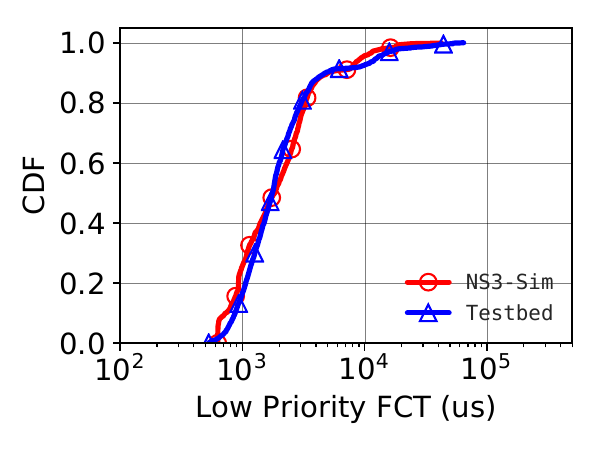}
            \caption{\emph{On-Off} - Med}
            \label{fig:TBVSSim-WLCOL-Med}
        \end{subfigure}%
        \begin{subfigure}{.32\linewidth}
            \includegraphics[width=\linewidth]{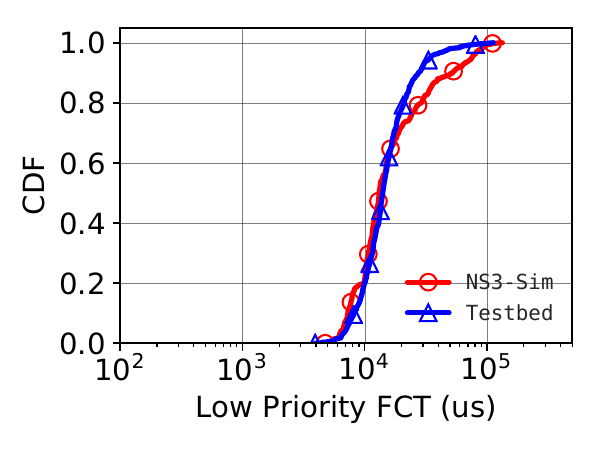}
            \caption{\emph{On-Off} - Long}
            \label{fig:TBVSSim-WLCOL-Long}
        \end{subfigure}%
    \end{minipage}%
    
    \begin{minipage}{\linewidth}
        \centering
        \begin{subfigure}{.32\linewidth}
            \includegraphics[width=\linewidth]{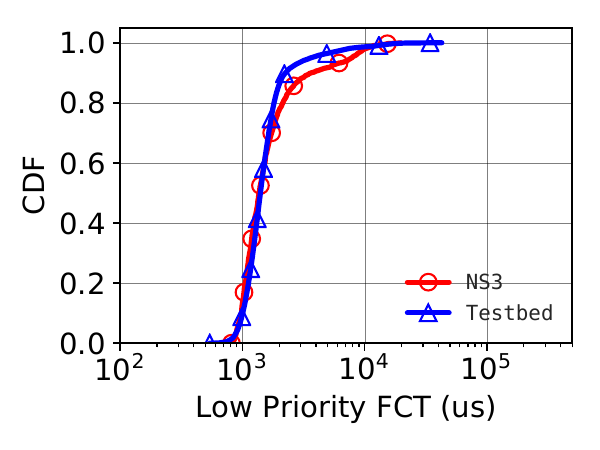}
            \caption{DAS - 128KB}
            \label{fig:TBVSSim-DAS-128KB}
        \end{subfigure}%
        \begin{subfigure}{.32\linewidth}
            \includegraphics[width=\linewidth]{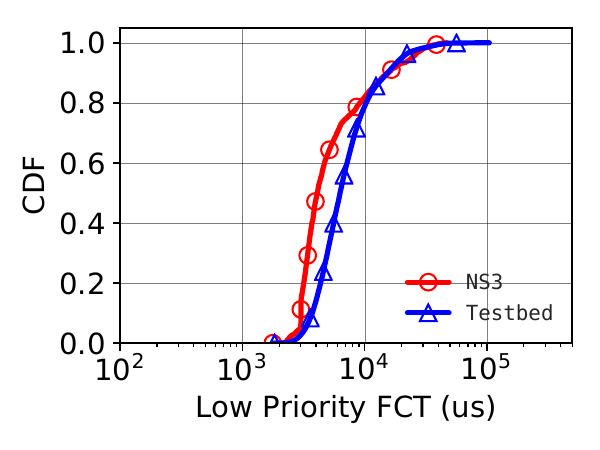}
            \caption{DAS - 1MB}
            \label{fig:TBVSSim-DAS-1MB}
        \end{subfigure}%
    \end{minipage}
    \caption{Compares performance of low-priority flows from testbed results and NS3 based simulation results under network scheduling~\S\ref{sec:eval-netsched} (a-c) and workload co-location~\S\ref{sec:eval-wl-col} (d-e) use-cases to establish the fidelity of our simulations.} 
    \label{fig:eval-TBVSSim}    
\end{figure}
\subsection{Testbed Evaluation}
\label{subsec:testbed-eval}
To establish the fidelity of our simulation results, we now repeat a set of experiment on a real testbed and compare the results with the simulation results.

\paragraph{Testbed Setup:}
For our testbed evaluations we provisioned 9 Dell servers on CloudLab~\cite{cloudlab} that are connected with a Mellanox SN2410 switch via 10Gbps links. Each server is powered by 10$\times$core Intel E5-2640v4 processor and 64GB RAM and runs on a linux based operating system. To enable network prioritization, we leverage linux queuing discipline at the end hosts -- a common practice employed in cloud systems -- whereas, inside the network, we enable strict priority queuing at the switch.
We focus on two use-cases of workload co-location~\S\ref{sec:eval-wl-col} and network scheduling~\S\ref{sec:eval-netsched} to draw the comparison between simulation and testbed results. Note that our testbed experiments lacks coverage of \emph{hybrid services} use-case because of practical constraints in realizing \nearopt{} in a real testbed. 
We configure the network switch and linux stack at the end-host using the same settings as used in our simulation with the exception of adjusting the \emph{RTO\textsubscript{min}} to 5ms -- the smallest value we could configure in our setup without using \emph{hrtimers}~\cite{hrtimer-sigcomm09}. We also repeat the NS3 experiments with the updated value of \emph{RTO\textsubscript{min}} for a fair comparison. We thoroughly examine the impact of this modification in~\S\ref{sec:micro}. 

Figure~\ref{fig:eval-TBVSSim} shows a general alignment between our testbed results and NS3 simulation results. However, a closer look reveals noticeable differences (e.g., at the tail for long flows in figure~\ref{fig:TBVSSim-WLCOL-Long}). We attribute these differences to fundamental disparities between NS3 and the Linux stack, particularly highlighting NS3's superior capability in enforcing strict prioritization at the end hosts. 
\section{Beyond Default Settings}
We now revisit default configurations that we used so far in our study.
Throughout this section, we focus on the workload co-location scenario (\S\ref{sec:eval-wl-col}) at a high load setting and fix the gradient update size to 4MB.
\label{sec:micro}

\begin{figure}[t]
  \begin{minipage}[t]{.48\columnwidth}
    \includegraphics[width=\columnwidth]{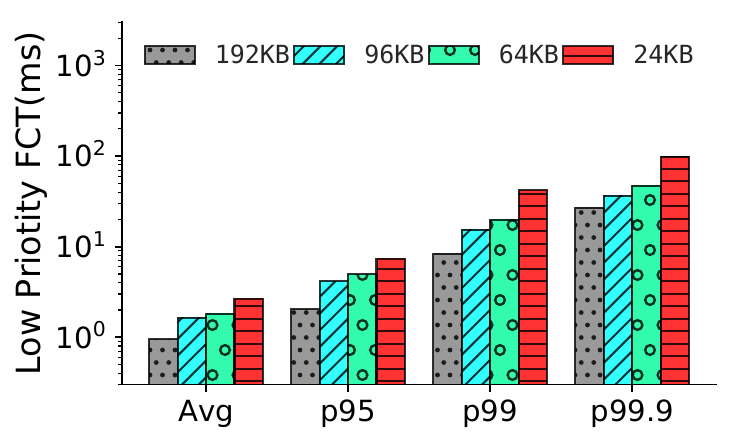}
    \caption{Shows that a smaller buffer size, a common case with low-priority queue, contributes to the performance degradation of TCP.}
	\label{fig:micro-buff}
  \end{minipage}
  \hfill
  \begin{minipage}[t]{.48\columnwidth}
    \includegraphics[width=\columnwidth]{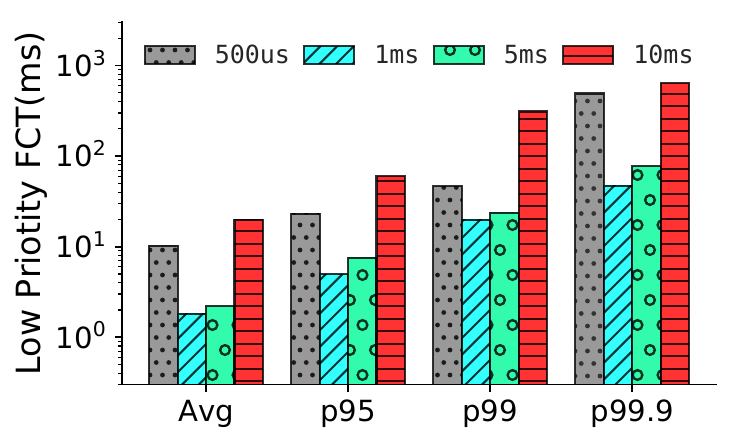}
    \caption{Shows that a too small or too high value of \emph{RTO\textsubscript{min}} severely degrades the performance of TCP for low-priority flows.}
	\label{fig:micro-rto-fct}
  \end{minipage}
\end{figure}

\paragraph{\textbf{Role of Buffer Sizes:}}
Data center network link speeds are increasing from 10Gbps to 40Gbps and 100Gbps however, the relative switch buffer sizes are shrinking. Studies show that per port per Gbps buffer size has shrunk from 192KB for 10Gbps to less than 10KB for 40Gbps switches~\cite{one-more-config}. Moreover, the available buffer size for low-priority queues is even smaller~\cite{fb-buffalloc}.
These settings can have detrimental impact on the performance of TCP for low-priority flows.
Following the \textit{Broadcom Tridant+} switch configuration~\cite{one-more-config} -- commonly deployed in data center networks, we assume per port buffer size of 192KB. We evaluate 4 different configuration of buffer sharing: 192KB indicates the availability of the entire port buffer for low-priority flow, 96KB represents the equal partitioning of port buffer between high and low-priority queues, 64KB refers to the default case in our experiments discuss in~\S\ref{sec:eval-main}, and 24KB represents the equal sharing of port buffer across 8 queues (typically 8 queues per port are available in switches).

Figure~\ref{fig:micro-buff} shows as the available buffer for low-priority queue shrinks, the performance of TCP deteriorates. 
For example, at p99 TCP's performance is 5.7$\times$ worse for 24KB case compared to 192KB case when entire port buffer is available to low-priority queue.
While a low buffer size is generally bad for TCP (i.e., fair-queuing case), this issue becomes even more pronounced when dealing with low-priority flows.

\paragraph{\textbf{Role of \textit{RTO\textsubscript{min}}:}}
The retransmission timeout (\emph{RTO}) indicates packet loss and triggers the packet retransmission. This value is estimated by continuously monitoring a flows RTT and incorporates \emph{RTO\textsubscript{min}} as a lower bound for on the \emph{RTO} value. Thus, it plays an important role in TCP's loss recovery process. Existing literature, on the other hand, argues in the favor of choosing a smaller value of RTO\textsubscript{min} for data center network in fairshare settings~\cite{hrtimer-sigcomm09,pfabric,fastlane,annulus,ndp}. Since, low-priority flows are prone to spurious timeouts (discussed in ~\S\ref{sec:motive}), one can argue that a lower value of \emph{RTO\textsubscript{min}} can exacerbate this issue. Overall, coming up with a right timeout value is a challenge in priority queue settings~\cite{aeolus}.

To investigate this in detail, we choose four different values of \emph{RTO\textsubscript{min}} commonly used in the literature: 500$\mu$s (following the 3$\times$ Avg RTT), 1ms (our default case), 5ms, and 10ms~\cite{dctcp,annulus,ndp}.
Figure~\ref{fig:micro-rto-fct} shows that a too small and too high value of \textit{RTO\textsubscript{min}} severely affects the flow completion times. This is primarily due to the fact that a larger value significantly contributes to latency in loss recovery, which becomes a major concern, particularly when dealing with tail losses where fast recovery is not an option. On the other hand, an excessively small value notably increases the occurrence of spurious timeouts.

\begin{figure}[t]
  \hfill
    \begin{minipage}[t]{.48\columnwidth}
	\includegraphics[width=\columnwidth]{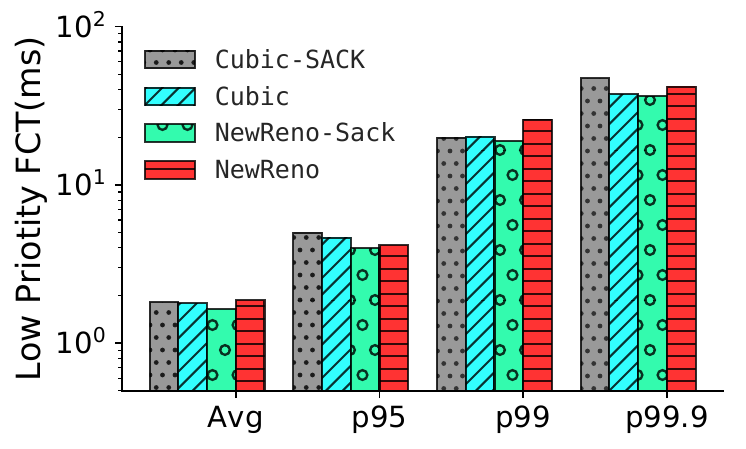}
    \caption{Compares the performance of various configurations of TCP. The choice of these configurations have negligible impact on the performance of TCP}
	\label{fig:micro-tcpvariants}
  \end{minipage}
  \hfill
  \begin{minipage}[t]{.48\columnwidth}
    \includegraphics[width=\columnwidth]{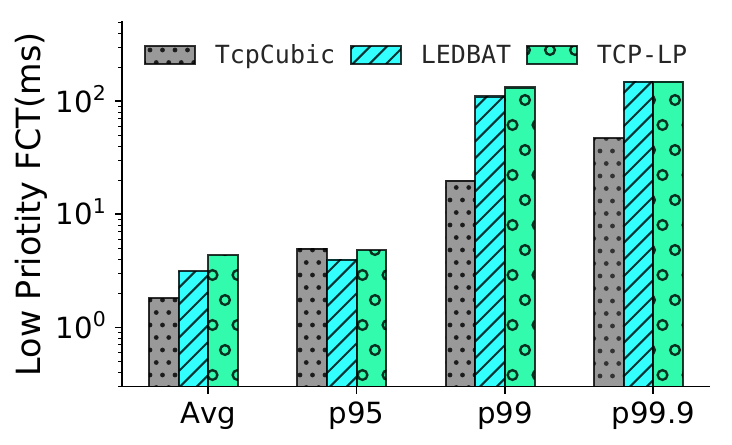}
    \caption{Shows that low-priority transport protocols designed for WAN settings are misfit for cloud environment.}
	\label{fig:micro-lpt}
  \end{minipage}  
\end{figure}

\paragraph{\textbf{Impact of TCP Configurations:}}
Our study focuses on TCP-Cubic with the SACK (Selective Acknowledgment) option enabled. TCP-Cubic is designed to maintain a plateau around \emph{WMAX}, which is determined after a packet loss event. However, in the presence of network prioritization which can cause excessive packet loss, TCP-Cubic, in theory, can act conservatively.
To evaluate this, we modify the TCP configurations by considering both TCP-Cubic and TCP-NewReno and comparing their performance with and without the SACK option enabled.
Figure~\ref{fig:micro-tcpvariants} shows that changing these configurations has minimal to no impact on TCP's performance.

\paragraph{\textbf{Low-Priority Transports for WAN:}}
Several proposals have looked at the need to optimize transport protocols for low-priority traffic under wide area network (WAN) settings~\cite{lpt-pcc-proteus,lpt-ledbat,lpt-tcp-lp}.
These proposals implement congestion control mechanisms specifically tailored for background traffic, which tend to be conservative to avoid disrupting higher priority traffic when network prioritization is absent.
However, due to their end-to-end nature, these protocols are susceptible to frequent stalls in their feedback loop. The challenges associated with such frequent stalls, combined with their conservative rate control, make them sub-optimal choices for low-priority flows under our scenario.

To evaluate this, we consider TCP-LP~\cite{lpt-tcp-lp} and LEDBAT~\cite{lpt-ledbat}. Both of these are delay based transport protocols designed for low-priority traffic for WAN. LEDBAT aims to maintain the bottleneck queuing delay around a specified \emph{target}. Estimating this bottleneck queuing delay for low-priority packets is challenging because it is influenced by both the queue lengths and the \emph{priority-induced} delay. To simplify this choice, we made an observation in our experiments that a low-priority packet, once enqueued, experiences the \emph{priority-induced} delay due to a single gradient update. This observation allowed us to approximate the \emph{target} value, which we set to 3.2ms. Figure~\ref{fig:micro-lpt} shows that both these protocols experience a significant increase in the FCTs. Specifically, compared to TCP, we observed up to 2.4$\times$ and 6.67$\times$ increase in FCTs on avg and at p99 respectively.
\section{{Improving TCP's Performance for Low-Priority Flows}}
\label{sec:future}
Our evaluation points out that for certain use-cases, TCP can experience severe performance degradation for low-priority flows. We discuss that this is primarily because of the transient starvation of low-priority queue. We now discuss (and do a preliminary evaluation) of two simple strategies for addressing these issues. 

\paragraph{\textbf{Weighted Fair Queuing instead of Priority Queues:}}
One approach to address this issue is to replace strict priority queuing (SPQ) with weighted fair queuing (WFQ). WFQ allows for differential service while avoiding starvation in low-priority queues. Recent work~\cite{aequitas} has highlighted the performance benefits of WFQ over SPQ, particularly in achieving service-level objectives (SLO) for remote procedural calls (RPC). By assigning a small weight to the lower priority class, WFQ ensures that low-priority flows receive a minimal fraction of the link share, thereby preventing stalls in their feedback loop while maintaining desired isolation across priority queues. 

\begin{figure*}[t]
    \centering
     \subfloat[Small Flows ]{\includegraphics[width=0.33\textwidth]{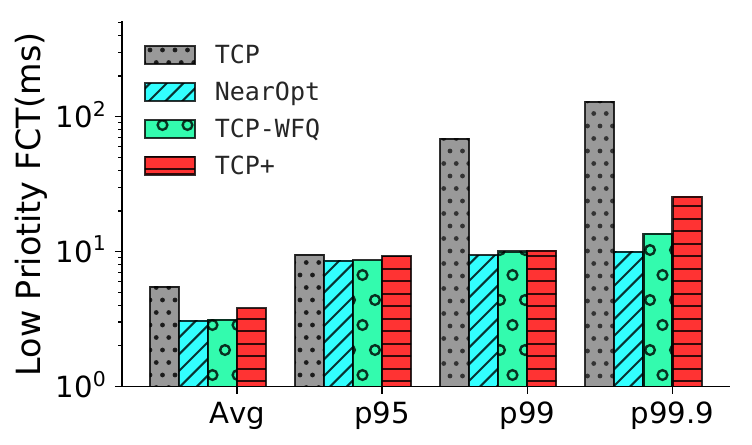}\label{fig:future-small}}
     \subfloat[Medium Flows]{\includegraphics[width=0.33\textwidth]{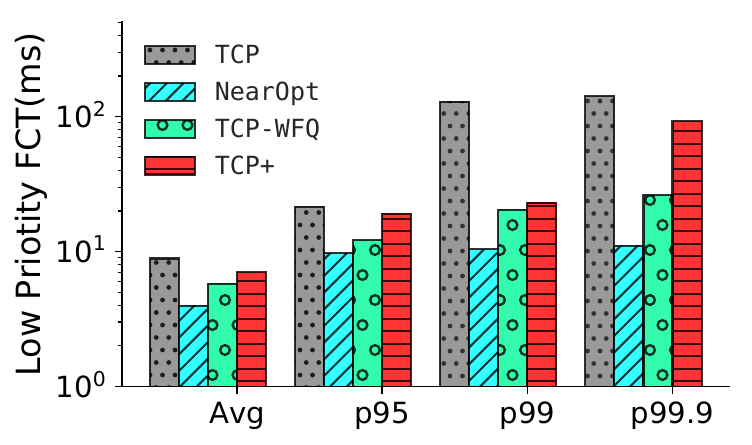}\label{fig:future-med}}
    \subfloat[Long Flows]{\includegraphics[width=0.33\textwidth]{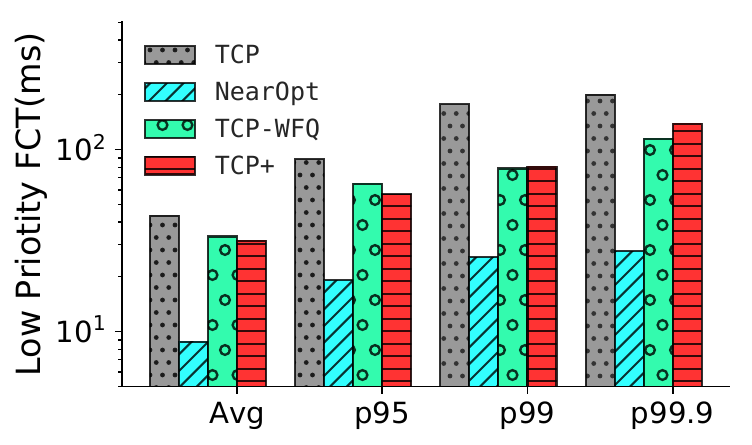}\label{fig:future-long}}
    \caption{Shows that efficacy of weighted fair queuing (WFQ) and TCP+. Both these approaches have the potential to limit the performance implications of network prioritization. }
\label{fig:eval-future}
\vspace{-0.1in}
\end{figure*}

To assess the advantages of using WFQ for low-priority flows and examine the impact on the high priority workload, we repeat workload co-location experiment from section~\S\ref{sec:eval-wl-col}. We choose the gradient update size of 12MB -- the case where TCP observed maximum performance degradation -- and replace SPQ with WFQ at the switches. We choose a weight distribution of 99:1 for high and low-priority classes respectively. 
Figure~\ref{fig:eval-future} shows that WFQ significantly improves TCP's performance. Specifically, at p99.9, we observe up to 9$\times$ and 5.4$\times$ reduction in the FCTs of small and medium flows respectively. However, our analysis revealed that WFQ resulted in an increase of 6\% in FCTs for high priority flows at p99.9.

While weighted fair queuing (WFQ) shows promise in enhancing TCP's performance for low-priority flows, its use presents a challenge in selecting the appropriate weight value for lower priority classes -- using a small value may render WFQ ineffective, while a large value can potentially interfere with the performance of high priority traffic. Moreover, when dealing with multiple lower priority classes~\cite{auto,pias}, additional caution is necessary in assigning minimum weights, warranting further investigation.

\paragraph{\textbf{Cross-Queue Congestion Notification (CQCN)}}
For certain systems that necessitate strict isolation across priority classes (e.g., DAS~\cite{das}) or in scenarios where changing the switch scheduling algorithm may not be an option, we discuss an end-host based approach that we term as \emph{Cross-Queue Congestion Notification}. Drawing inspiration from DCTCP~\cite{dctcp}, we make a simple yet an important observation: employing Explicit Congestion Notification (ECN) in a \emph{cross-queue} manner enables TCP to discern \emph{priority-induced} congestion at the end-host. Such a cross-queue signal can enable TCP to make an informed decision to stop packet transmission and halt congestion control for low-priority flows upon experiencing \emph{priority-induced} congestion.
Consequently, CQCN can avoid overwhelming low-priority queues, effectively limiting packet losses and spurious congestion backoffs for low-priority flows.

To enable CQCN, low-priority flows can leverage \emph{low-overhead} yet \emph{high-priority} probe packets that act to create a secondary, \emph{non-blocking}, feedback loop between end-hosts. Similar to DCTCP, probe packets get marked if the length of the high priority queue exceeds a certain threshold -- indicating \emph{priority-induced} congestion.
However, unlike DCTCP -- which may encounter feedback stalls when employed for low-priority flows -- the secondary feedback loop continues to provide a low-priority TCP sender with updates about the state of high priority traffic. Cross-queue can leverage existing ECN support in network switches without requiring any modification inside the network.

To evaluate the potential benefits of such an approach we repeat the workload co-location experiment from section~\S\ref{sec:eval-wl-col} with gradient update size of 12MB. Figure~\ref{fig:eval-future} shows that such an approach, shown as TCP+, can enhance TCP's performance. Specifically, we observe up to 6$\times$ decrease in FCTs for medium flows at p99. 

While our preliminary evaluation indicates the promise of this approach, it also present several challenges. For instance, despite the probe packets being \emph{low-overhead}, per-flow probing can interfere with high priority traffic, thereby violating the isolation constraint. While probing overhead is inevitable with this approach, its impact can be minimized by multiplexing probing feedback across all flows between a pair of end-hosts.
Similarly, another challenge lies in selecting the appropriate probe marking threshold. A low threshold may inaccurately indicate excessive congestion in the high priority queue, while a high threshold may render the probing feedback ineffective, causing TCP to fail in responding to congestion in the high priority queue. 
Overall, these challenges present intriguing avenues for future research on this problem.
\section{Discussion} 
\label{sec:discuss}
\paragraph{\textbf{Emerging Data Center Transport Protocols:}}
In our study we focused on the performance evaluation of TCP for low-priority flows. But the challenges introduced by network prioritization can affect \emph{fair-transports}\footnote{End-host based transport protocols that are designed to fairly share the network bandwidth} in general. Majority of the existing data center transport protocols follow the fairshare policy and rely on an end-end feedback loop between the sender and the receiver~\cite{dctcp,swift,timely,dcqcn,hpcc}. We hypothesize that these proposals are susceptible to performance issues when used for low-priority flows however, the sensitivity of these protocols to network prioritization may vary depending on their choice of congestion signal (e.g., delay, ECN etc.,) and the congestion control algorithm~\cite{ecn-or-delay,on-ramp}. For example, a recent work~\cite{bfc} shows that despite HPCC~\cite{hpcc} leveraging a rich network state (e.g., queue size and link utilization information), its performance suffers under network prioritization. The increasing interest in the large-scale deployment of these proposals~\cite{hpcc,dcqcn} may require a separate study on their performance evaluation for low-priority flows.


\paragraph{\textbf{Dual Congestion Control for Data Center Applications:}}
Annulus~\cite{annulus}, propose a dual congestion control loop to improve the co-location of WAN and data center network traffic. Annulus relies on a direct congestion signal from network switches to the end host.
Their findings align with our observations, indicating that augmenting TCP with additional congestion information can improve TCP's performance. A recent work, On-Ramp~\cite{on-ramp}, advocates for the modularization of congestion control into two independent components to handle transient and persistent congestion separately. On-Ramp leverages one-way delay estimates to classify congestion. Similar to On-Ramp, we discuss that the use of a dual congestion control loop to enhance TCP's performance in the presence of network prioritization.


\paragraph{\textbf{Reconfigurable Data Center Networks (RDCN)}:} 
Under RDCNs, the path between the end hosts may alternate rapidly between slow electrical network and fast optical network resulting in an order of magnitude difference in latency and throughput. Chen et. al.~\cite{rdcn-tdtcp} shows that existing TCP variants are unable to take full advantage of the additional capacity provided by RDCNs. Our findings (see 
~\S\ref{fig:eval-wl-col}) also highlight that TCP suffers significantly under high fluctuation in the available network bandwidth.

\paragraph{\textbf{Switch Buffer Sharing Algorithms:}}
In~\S\ref{sec:micro}, we evaluate the impact of buffer size on the performance of TCP for low-priority flows. We consider static partitioning of the available buffer between the queues. However, the switch buffer is often dynamically shared across queues~\cite{dt,abm}. Buffer sharing algorithms provides control over queue lengths across priority classes via a configurable parameter $\alpha$. Typically, a high (resp.  low) value of $\alpha$ is selected for high (resp. low) priority classes. A recent work~\cite{abm} shows that under a large number of priority classes (typically 8), the dynamic sharing can lead to starvation for lower priority classes. We observe similar behavior when the available buffer size for low-priority queues is set to a low value.

\paragraph{\textbf{End-host Stack Delays:}}
Feedback delay for an end-host based transport (e.g., TCP) can broadly be divided in two categories: 1) end-host stack delays, 2) fabric delays.
In this study, we focused on the fabric delays caused by network prioritization. A recent work~\cite{stack-delays} showed that stack induced delays can be very large especially for networks with high link speeds. We believe that stack induced delays will further deteriorate TCP's problem for the scenarios considered in this study. 
\section{Conclusion} 
\label{sec:conclusion}
In this study we show that the use of network prioritization in a cloud system can pose serious challenges to the use of TCP for low-priority flows. Our study thoroughly investigates the resulting performance implications under a range of workloads, load settings and a variety of configurations. We conclude that TCP's performance is severely impacted under workloads that exhibits \emph{on-off} behavior on the network. We demonstrate that TCP's performance can be improved by incorporating the state of high priority traffic in the congestion control of low-priority flows which opens up exciting an avenue for future research.
\newpage
\bibliographystyle{splncs04}
\bibliography{ref}
\end{document}